\documentclass[pre,twocolumn,showpacs]{revtex4}
\usepackage{epsfig}

\begin{document}

\title{Quasi-planar steep water waves}
\author{V.~P. Ruban}
\email{ruban@itp.ac.ru}
\affiliation{Landau Institute for Theoretical Physics,
2 Kosygin Street, 119334 Moscow, Russia} 

\date{\today}

\begin{abstract}
A new description for highly nonlinear potential water waves is 
suggested, where weak 3D effects are included as small corrections to  
exact 2D equations written in conformal variables.
Contrary to the traditional approach, a small parameter in this theory is not 
the surface slope, but it is the ratio of a typical wave length to a large 
transversal scale along the second horizontal coordinate. A first-order 
correction for the Hamiltonian functional is calculated, and the corresponding
equations of motion are derived for steep water waves 
over an arbitrary inhomogeneous quasi-1D bottom profile.

\end{abstract}

%47.10.+g General theory
%47.15.Hg Potential flows
%47.35.+i Hydrodynamic waves
\maketitle

%%%%%%%%%%%%%%%%%%%%%%%%%%%%%%%%%%%%%%%%%%%%%%%%%%%%%%%%%%%%%%%%%%%%%%%%%%%

\section{Introduction}

The problem of water waves is one of the classical fields of the hydrodynamics,
and it has been studied extensively over many years.
Starting from the middle of 90-s, in the theory of two-dimensional (2D) 
potential flows of an ideal fluid with a free surface, the so called conformal
variables have been actively employed \cite{DKSZ96,DZK96,DLZ95,ZD96,Lvov97}. 
With these variables, highly nonlinear equations of motion for planar 
water waves can be written 
in an exact and compact form containing integral operators diagonal in the
Fourier representation. Such integrodifferential equations are very 
suitable for numerical simulation, because effective computer programs for the
discrete fast Fourier transform (FFT) are now available
(see, e.g., \cite{fftw3}). Based on these equations, a significant progress 
has been achieved in the study of nonlinear dynamics of water waves, 
including the mechanism of sudden formation of the giant waves \cite{D2001,ZDV2002}. 
Recently, the exact 2D description has been 
generalized to the case of a highly space- and time-inhomogeneous bottom 
profile \cite{R2004PRE,physics/0411011}. However, the real water waves are 
never ideally two-dimensional. Therefore there is a need of a theory, which
could describe strongly nonlinear, even breaking waves and, from the other hand,
it would take into account  3D effects, at least as weak corrections. 
In present work such a highly nonlinear weakly 3D theory is suggested 
as an extension of the exact 2D theory.
It should be emphasized that existing approximate nonlinear evolution equations
for water waves, for example the famous Kadomtsev-Petviashvily equation,
equations of Boussinesq type \cite{Boussinesq}, or the equations obtained 
by Matsuno \cite{Matsuno}, are valid just for weakly nonlinear  
water waves, but not for overturning or breaking waves.
 
It is a well known fact that a very significant difficulty in the 3D theory
of potential water waves is the general impossibility to solve the Laplace
equation for the velocity potential $\varphi(x,y,q,t)$,
\begin{equation}\label{Laplace_xyq}
\varphi_{xx}+\varphi_{yy}+\varphi_{qq}=0,
\end{equation}
in the flow region $-H(x,q)\le y\le \eta(x,q,t)$ between the 
(static for simplicity) bottom and a
time-dependent free surface, with the given boundary conditions
\begin{equation}\label{boundary_cond_Laplace_xyq}
\varphi|_{y=\eta(x,q,t)}= \psi(x,q,t),\quad 
({\partial\varphi}/{\partial n})|_{y=-H(x,q)}=0.
\end{equation}
(Here $x$ and $q$ are the horizontal Cartesian coordinates, $y$ is the vertical
coordinate, while the symbol $z$ will be used for the complex combination
$z=x+iy$). Therefore a compact expression is absent for the Hamiltonian
functional of the system,
\begin{eqnarray}
{\cal H}\{\eta,\psi\}&=&\frac{1}{2}\int dx\, dq
\int\limits_{-H(x,q)}^{\eta(x,q,t)}(\varphi_x^2+\varphi_y^2+\varphi_q^2)dy
\nonumber\\
&&+\frac{g}{2}\int\eta^2dx\,dq\equiv {\cal K}\{\eta,\psi\}
+{\cal P}\{\eta\},
\label{Hamiltonian_eta_psi}
\end{eqnarray} 
(the sum of the kinetic energy of the fluid and the potential energy in
the vertical gravitational field $g$). The Hamiltonian determines the canonical 
equations of motion (see \cite{Z1999,RR2003PRE,DKZ2004}, 
and references therein)
\begin{equation}\label{Hamiltonian_equations_eta_psi}
 \eta_t=\frac{\delta{\cal H}}{\delta\psi},\qquad
-\psi_t=\frac{\delta{\cal H}}{\delta\eta}
\end{equation}
in accordance with the variational principle $\delta\int \tilde{\cal L}dt=0$,
where the Lagrangian is
\begin{equation}\label{Lagrangian_eta_psi}
\tilde{\cal L}=\int\psi\eta_t \,dx\,dq-{\cal H}.
\end{equation}

In the traditional approach, the problem is partly solved by an asymptotic 
expansion of the kinetic energy ${\cal K}$ on a small parameter --- the slope of the
surface (see \cite{Z1999,DKZ2004}, and references therein). As the result,
a weakly nonlinear theory arises, which is not good to describe 
large-amplitude waves (see \cite{LZ2004} for a discussion about the limits of 
such theory). The theory developed in present work is based on another small
parameter --- the ratio of a typical length of the waves propagating along the
$x$-axis, to a large scale along the transversal horizontal direction,
denoted by $q$. Thus, we define $\epsilon=(l_x/l_q)^2\ll 1$ and note: 
the less this parameter, the less our flow differs from a purely 2D flow.
The profile $y=\eta(x,q,t)$ of the free surface, the boundary value of the
velocity potential $\psi(x,q,t)\equiv\varphi(x,\eta(x,q,t),q,t)$, and a given 
bottom profile $y=-H(x,q)$ are allowed to depend strongly on the coordinate $x$,
while the derivatives over the coordinate $q$ will be supposed small:
$|\eta_q|\sim\epsilon^{1/2}$, $|\psi_q|\sim\epsilon^{1/2}$, 
$|H_q|\sim\epsilon^{1/2}$.

The paper is organized as follows. Sec. II is devoted to a general description
of the present approach. In Sec. III, an explicit expression for the first-order
correction ${\cal K}^{(1)}$ is obtained, thus we can take into account, in the 
main approximation, weak 3D effects.

\section{General idea of the method}

In the same manner as in the exact 2D theory \cite{R2004PRE,physics/0411011},
instead of the Cartesian coordinates $x$ and $y$, we use curvilinear conformal
coordinates  $u$ and $v$, which make the free surface and the bottom
effectively flat:
\begin{equation}\label{conformal_mapping}
x+iy\equiv z=z(u+iv,q,t),\quad -\infty <u<+\infty,\quad 0\le v\le 1,
\end{equation}
where $z(w,q,t)$ is an analytical on the complex variable $w\equiv u+iv$
function without any singularities in the flow region $0\le v\le 1$.
Now the bottom corresponds to $v=0$, while on the free surface $v=1$.
The boundary value of the velocity potential is 
$\varphi|_{v=1}\equiv \psi(u,q,t)$. In the case of a non-horizontal curved
bottom, it is convenient to represent the conformal mapping $z(w,q,t)$ as 
a composition of two conformal mappings $w\mapsto\zeta\mapsto z$, 
similarly to works \cite{R2004PRE,physics/0411011}:
\begin{equation}\label{Z_zeta_w}
z(w,q,t)=Z(\zeta(w,q,t),q).
\end{equation}
Here the intermediate function $\zeta(w,q,t)$ possesses the property 
$\mbox{Im\,}\zeta(u+0i,q,t)=0$, thus resulting in the important relation 
\begin{equation}\label{xi}
\zeta(u+i,q,t)\equiv\xi(u,q,t)=(1+i\hat R)\rho(u,q,t),
\end{equation}
where $\rho(u,q,t)$ is a purely real function, and $\hat R=i\tanh\hat k$ 
(here $\hat k\equiv -i\hat\partial_u$) is the anti-Hermitian operator,
 which is diagonal in the Fourier representation:
it multiplies the Fourier-harmonics 
$\rho_k(q,t)\equiv\int\rho(u,q,t)e^{-iku}du$ by $R_k=i\tanh k$,  
so that
\begin{eqnarray}
\hat R \rho(u,q,t)&=&\int [i\tanh k] \rho_k(q,t)e^{iku}\frac{dk}{2\pi}
\nonumber\\
&=&\mbox{P.V.}\int\frac{\rho(\tilde u,q,t)\,d\tilde u}
{2\sinh[(\pi/2)(\tilde u-u)]}.
\label{R_def} 
\end{eqnarray}
(P.V. means the principal value integral.)
A known analytical function $Z(\zeta,q)$ determines parametrically 
the static bottom profile:
\begin{equation}\label{bottom_profile}
X^{[b]}(r,q)+iY^{[b]}(r,q)=Z(r,q),
\end{equation}
where $r$ is a real parameter running from $-\infty$ to $+\infty$.
The profile of the free surface is now given (in the parametric form as well) 
by the formula
\begin{equation}\label{surface_profile}
X^{[s]}(u,q,t)+iY^{[s]}(u,q,t)\equiv Z^{[s]}(u,q,t)=Z(\xi(u,q,t),q).
\end{equation}
For equations to be more short, below we do not indicate the arguments $(u,q,t)$
of the functions $\psi$, $\xi$ É $\bar\xi$ (the overline denotes the complex
conjugate). Also, we introduce the notation $Z'(\xi)\equiv \partial_\xi Z(\xi,q)$.
The Lagrangian of the system in the variables $\psi$, $\xi$, and $\bar\xi$ can
be re-written as follows (compare to \cite{R2004PRE}):
\begin{eqnarray}
{\cal L}&=&\int Z'(\xi)\bar Z'(\bar\xi)\left[
\frac{\xi_t\bar \xi_u -\bar\xi_t\xi_u}{2i}\right]\psi \,du\,dq \nonumber\\
&-&{\cal K}\{\psi,Z(\xi),\bar Z(\bar\xi)\}\nonumber\\
&-&\frac{g}{2}\int\left[\frac{Z(\xi)-\bar Z(\bar\xi)}{2i}\right]^2
\left[\frac{Z'(\xi)\xi_u+\bar Z'(\bar\xi)\bar\xi_u}{2}\right]du\,dq
\nonumber\\
\label{Lagrangian_Z_psi}
&+&\int\Lambda\left[\frac{\xi-\bar\xi}{2i}
-\hat R \left(\frac{\xi+\bar\xi}{2}\right)\right]du\,dq,
\end{eqnarray}
where the indefinite real Lagrangian multiplier $\Lambda(u,q,t)$ has been
introduced in order to take into account the relation (\ref{xi}).
Equations of motion follow from the variational principle 
$\delta{\cal A}=0$, with the action ${\cal A}\equiv\int {\cal L}dt$.
So, the variation by $\delta\psi$ gives us the first equation of motion ---
the kinematic condition on the free surface:
\begin{equation}\label{kinematic_1}
|Z'(\xi)|^2\,\mbox{Im\,}(\xi_t\bar\xi_u)=\frac{\delta{\cal K}}{\delta\psi}.
\end{equation}
Let us divide this equation by $|Z'(\xi)|^2|\xi_u|^2$ and use the analytical 
properties of the function $\xi_t/\xi_u$. As the result, we obtain the
time-derivative-resolved equation
\begin{equation}\label{kinematic}
\xi_t=\xi_u(\hat T+i)\left[\frac{(\delta{\cal K}/\delta\psi)}
{|Z'(\xi)|^2|\xi_u|^2}\right],
\end{equation}
where the linear operator $\hat T\equiv\hat R^{-1}=-i\coth \hat k$ 
has been introduced. Further, the variation of the action ${\cal A}$ by 
$\delta\xi$ gives us the second equation of motion:
\begin{eqnarray}
&&\left[\frac{\psi_u \bar\xi_t-\psi_t\bar\xi_u}{2i}\right]|Z'(\xi)|^2=
\left(\frac{\delta{\cal K}}{\delta Z}\right)Z'(\xi)
\nonumber\\
&&+\frac{g}{2i}\,\mbox{Im}\Big(Z(\xi)\Big)|Z'(\xi)|^2\bar\xi_u 
-\frac{(1+i\hat R)\Lambda}{2i}.
\label{dynamic_1}
\end{eqnarray}
After multiplying Eq.(\ref{dynamic_1}) by $-2i\xi_u$ we have
\begin{eqnarray}
&&\left\{[\psi_t+g\,\mbox{Im\,}Z(\xi)]|\xi_u|^2-\psi_u \bar\xi_t\xi_u
\right\}|Z'(\xi)|^2\nonumber\\
&&=(1+i\hat R)\tilde\Lambda
-2i\left(\frac{\delta{\cal K}}{\delta Z}\right)Z'(\xi)\xi_u,
\label{dynamic_2}
\end{eqnarray}
where $\tilde\Lambda$ is another real function. Taking the imaginary part of
Eq.(\ref{dynamic_2}) and using Eq.(\ref{kinematic_1}), we find 
$\tilde\Lambda$:
\begin{equation}\label{tilde_Lambda}
\tilde\Lambda =\hat T\left[\psi_u \frac{\delta{\cal K}}{\delta\psi}\right]
+2\hat T\,\mbox{Re}\left[
\left(\frac{\delta{\cal K}}{\delta Z}\right)Z'(\xi)\xi_u\right].
\end{equation}
After that, the real part of Eq.(\ref{dynamic_2}) gives us the  Bernoulli
equation in a general form:
\begin{eqnarray}
\psi_t+g\,\mbox{Im\,}Z(\xi)&=&
\psi_u\hat T\left[\frac{(\delta{\cal K}/\delta\psi)}
{|Z'(\xi)|^2|\xi_u|^2}\right]+
\frac{\hat T\left[\psi_u ({\delta{\cal K}}/{\delta\psi})\right]}
{|Z'(\xi)|^2|\xi_u|^2}\nonumber\\
\label{Bernoulli}
&+&\frac{2\,\mbox{Re}\left((\hat T-i)
\left[({\delta{\cal K}}/{\delta Z})Z'(\xi)\xi_u\right]\right)}
{|Z'(\xi)|^2|\xi_u|^2}.
\end{eqnarray}

Equations (\ref{kinematic}) and (\ref{Bernoulli}) completely determine
the evolution of the system, provided the kinetic energy functional 
${\cal K}\{\psi,Z,\bar Z\}$ is explicitly given. It should be emphasized that
in our description a general expression for ${\cal K}$ remains unknown.
However, under the conditions $|z_q|\ll 1$, $|\varphi_q|\ll 1$, the potential
$\varphi(u,v,q,t)$ is efficiently expanded into a series on the powers of the
small parameter $\epsilon$:
\begin{equation}\label{varphi_expansion}
\varphi=\varphi^{(0)} +\varphi^{(1)}+\varphi^{(2)}+\dots,\qquad 
\varphi^{(n)}\sim \epsilon^n,
\end{equation}
where $\varphi^{(n+1)}$ can be calculated from $\varphi^{(n)}$, 
and the zeroth-order term $\varphi^{(0)}=\mbox{Re\,}\phi(w,q,t)$ 
is the real part of an easily represented (in integral form) 
analytical function with the boundary conditions
$\mbox{Re\,}\phi|_{v=1}=\psi(u,q,t)$, $\mbox{Im\,}\phi|_{v=0}=0$.
Correspondingly, the kinetic energy functional will be written in the form
\begin{equation}\label{H_expansion}
{\cal K}={\cal K}^{(0)}+{\cal K}^{(1)}+{\cal K}^{(2)}+\dots,\qquad 
{\cal K}^{(n)}\sim \epsilon^n,
\end{equation}
where ${\cal K}^{(0)}\{\psi\}$ is the kinetic energy of a purely 2D flow,
\begin{eqnarray} 
{\cal K}^{(0)}\{\psi\}&=&\frac{1}{2}\int 
[(\varphi^{(0)}_u)^2+(\varphi^{(0)}_v)^2]\,du\,dv\,dq\nonumber\\
&=&-\frac{1}{2}\int \psi\hat R \psi_u \,du\,dq,
\label{K_0} 
\end{eqnarray}
and other terms are corrections due to gradients along $q$. Now we are going to
calculate the first-order correction ${\cal K}^{(1)}$.

\section{The first-order corrections}

As the result of the conformal change of two variables, the kinetic energy
functional is determined by the expression
\begin{equation}\label{H_full}
{\cal K}=\frac{1}{2}\int\left[\varphi_u^2+\varphi_v^2
+J({\bf Q}\cdot\nabla\varphi)^2\right]du\, dv \,dq,
\end{equation}
where the conditions $x_u=y_v$, $x_v=-y_u$ have been taken into account, 
and the following notations are used:
$$
J\equiv|z_u|^2, \qquad 
({\bf Q}\cdot\nabla\varphi)\equiv a\varphi_u+b \varphi_v+\varphi_q,
$$
$$
 a=\frac{x_v y_q-x_q y_v}{J}\sim\epsilon^{1/2}, 
\quad b=\frac{y_u x_q-y_q x_u}{J}\sim\epsilon^{1/2}.
$$
Consequently, the Laplace equation in the new coordinates takes the form
\begin{equation}\label{Laplace_uvq}
\varphi_{uu}+\varphi_{vv}
+\nabla\cdot({\bf Q}J({\bf Q}\cdot\nabla\varphi))=0,
\end{equation}
with the boundary conditions
\begin{equation}
\varphi|_{v=1}=\psi(u,q,t),\quad
[\varphi_v+bJ(\varphi_q+a\varphi_u+b \varphi_v)]|_{v=0}=0.
\end{equation}
Under the condition $\epsilon\ll 1$ it is possible to write the solution as the
series (\ref{varphi_expansion}), with the zeroth-order term satisfying
the 2D Laplace equation
$$
\varphi_{uu}^{(0)}+\varphi_{vv}^{(0)}=0,
\qquad \varphi|_{v=1}=\psi(u,q,t),\qquad \varphi_v|_{v=0}=0.
$$
Thus, it can be represented as $\varphi^{(0)}=\mbox{Re\,}\phi(w,q,t)$, where
\begin{equation}\label{phi}
\phi(w,q,t)=\int\frac{\psi_k(q,t)e^{ikw}}{\cosh k}\frac {dk}{2\pi},
\end{equation}
\begin{equation}
\psi_k(q,t)\equiv\int\psi(u,q,t)e^{-iku}du.
\end{equation}
On the free surface
\begin{equation}\label{Psi_def}
\phi(u+i,q,t)\equiv\Psi(u,q,t)=(1+i\hat R)\psi(u,q,t).
\end{equation}
For all the other terms in Eq.(\ref{varphi_expansion}) we have the relations
\begin{equation}\label{recur}
\varphi_{uu}^{(n+1)}+\varphi_{vv}^{(n+1)}
+\nabla\cdot({\bf Q}J({\bf Q}\cdot\nabla\varphi^{(n)}))=0
\end{equation}
and the boundary conditions $\varphi^{(n+1)}|_{v=1}=0$,
$$
[\varphi_v^{(n+1)}+bJ(\varphi_q^{(n)}+a\varphi_u^{(n)}
+b \varphi_v^{(n)})]|_{v=0}=0.
$$
Noting that 
$\int(\varphi_u^{(0)}\varphi_u^{(1)}+\varphi_v^{(0)}\varphi_v^{(1)})\,du\,dv=0$
(it is easily seen after integration by parts), we have in the first
approximation
\begin{eqnarray}
{\cal K}^{(1)}&=&\frac{1}{2}\int
J(\varphi_q^{(0)}+a\varphi_u^{(0)}+b \varphi_v^{(0)})^2du\, dv \,dq
\nonumber\\
&=&\frac{1}{2}\int
z_u\bar z_u\left[\mbox{Re}\left(\phi_q-\frac{\phi_u z_q}{z_u}\right)
\right]^2du \,dv\, dq.
\label{K1a}
\end{eqnarray}
Since  $z(w)$ and $\phi(w)$ are represented as
$z(u+iv)=e^{\hat k(1-v)}Z^{[s]}(u)$ and $\phi(u+iv)=e^{\hat k(1-v)}\Psi(u)$, 
we can use for $v$-integration the following formulas:
\begin{eqnarray}
&&\int du\int_0^1 [e^{\hat k(1-v)}A(u)]
\overline{[e^{\hat k(1-v)}B(u)]}\,dv\nonumber\\
&=&\int \left(\frac{e^{2k}-1}{2k}\right) A_k\overline{B_k}\frac{dk}{2\pi}
\nonumber\\
&=&-\frac{i}{2}\int \overline{B(u)}\,\hat\partial_u^{-1}A(u)\, \,du
\nonumber\\
&&+\frac{i}{2}\int \overline{B^{[b]}(u)}\,\hat\partial_u^{-1}A^{[b]}(u)\,du, 
\end{eqnarray}
with $A^{[b]}(u)=e^{\hat k}A(u)$, $B^{[b]}(u)=e^{\hat k}B(u)$. As the result, 
we obtain from Eq.(\ref{K1a}) the expression of the form
${\cal K}^{(1)}={\cal K}^{(1)}_{[s]}-{\cal K}^{(1)}_{[b]}$, where
${\cal K}^{(1)}_{[s]}={\cal F}\{\Psi,\overline{\Psi},Z,\overline{Z}\}$,
${\cal K}^{(1)}_{[b]}={\cal F}\{\Psi^{[b]},\overline{\Psi^{[b]}},Z^{[b]},
\overline{Z^{[b]}}\}$, with $Z=Z^{[s]}$, $Z^{[b]}=e^{\hat k}Z$, 
$\Psi^{[b]}=e^{\hat k}\Psi=[\cosh\hat k]^{-1}\psi$. The functional
${\cal F}$ is defined below:
\begin{eqnarray}
{\cal F}&=&\frac{i}{8}
\int(Z_u\Psi_q-Z_q\Psi_u)\partial_u^{-1}
\overline{(Z_u\Psi_q-Z_q\Psi_u)}\,du\,dq
\nonumber\\
&+&\frac{i}{16}\int\Bigg\{\left[
(Z_u\Psi_q-Z_q\Psi_u)^2/{Z_u}\right]\overline{Z} 
\nonumber\\
&&\qquad - Z\,\overline{\left[(Z_u\Psi_q-Z_q\Psi_u)^2/{Z_u}\right]}
\Bigg\}\,du\,dq.
\label{K_1_s}
\end{eqnarray}
From here one can express the variational derivatives 
$(\delta{\cal K}^{(1)}/{\delta\psi})$ and $(\delta{\cal K}^{(1)}/{\delta Z})$ 
by the formulas
\begin{eqnarray}
\frac{\delta{\cal K}^{(1)}}{\delta\psi}&=&\left[
(1-i\hat R)\frac{\delta{\cal F}}{\delta\Psi}
+(1+i\hat R)\frac{\delta{\cal F}}{\delta\overline{\Psi}}
\right]\nonumber\\
&-&[\cosh\hat k]^{-1}\left(
\frac{\delta{\cal K}^{(1)}_{[b]}}{\delta\Psi^{[b]}}
+\frac{\delta{\cal K}^{(1)}_{[b]}}{\delta\overline{\Psi^{[b]}}}
\right),
\end{eqnarray}
\begin{equation}
\frac{\delta{\cal K}^{(1)}}{\delta Z}=
\frac{\delta{\cal F}}{\delta Z}-
e^{-\hat k}\left(\frac{\delta{\cal K}^{(1)}_{[b]}}
{\delta Z^{[b]}}\right).
\end{equation}
The derivatives
$(\delta{\cal F}/{\delta\Psi})$, $(\delta{\cal F}/{\delta Z})$,
$(\delta{\cal K}^{(1)}_{[b]}/{\delta\Psi^{[b]}})$, 
and $(\delta{\cal K}^{(1)}_{[b]}/{\delta Z^{[b]}})$
are calculated in a standard manner, for instance,
\begin{eqnarray}
\frac{\delta{\cal F}}{\delta\Psi}=\frac{i}{8}\,
Z_q\,\left[\overline{(Z_u\Psi_q-Z_q\Psi_u)}+ 
\hat\partial_u[(\Psi_q-Z_q\Psi_u/{Z_u})\overline{Z}]\right]&&
\nonumber\\
-\frac{i}{8}\,
Z_u\,\hat\partial_q\left[
\hat\partial_u^{-1}\overline{(Z_u\Psi_q-Z_q\Psi_u)}+ 
(\Psi_q-Z_q\Psi_u/{Z_u})\overline{Z}\right],&&\nonumber
\end{eqnarray}
\begin{eqnarray}
\frac{\delta{\cal F}}{\delta Z}=-\frac{i}{8}\,
\Psi_q\,\left[\overline{(Z_u\Psi_q-Z_q\Psi_u)}+ 
\hat\partial_u[(\Psi_q-Z_q\Psi_u/{Z_u})\overline{Z}]\right]&&
\nonumber\\
+\frac{i}{8}\,
\Psi_u\,\hat\partial_q\left[
\hat\partial_u^{-1}\overline{(Z_u\Psi_q-Z_q\Psi_u)}+ 
(\Psi_q-Z_q\Psi_u/{Z_u})\overline{Z}\right]&&\nonumber\\
+\frac{i}{16}\,\left[\hat\partial_u[(\Psi_q-Z_q\Psi_u/Z_u)^2
\overline{Z}]
-\overline{(\Psi_q-Z_q\Psi_u/{Z_u})^2{Z_u}}
\right].&&\nonumber
\end{eqnarray}
Now one can  substitute $(\delta{\cal K}/\delta\psi)\approx -\hat R\psi_u
+(\delta{\cal K}^{(1)}/\delta\psi)$ and 
$(\delta{\cal K}/\delta Z)\approx(\delta{\cal K}^{(1)}/\delta Z)$
into the equations of motion (\ref{kinematic}) and (\ref{Bernoulli}),
keeping in mind that  $Z=Z(\xi, q)$, $Z_u= Z'(\xi)\xi_u$, 
$Z_q= Z'(\xi)\xi_q+\partial_q Z$, $Z^{[b]}=Z([\cosh\hat k]^{-1}\rho,q)$, 
and so on. Thus, the weakly 3D equations of motion are completely derived,
and our main goal is achieved.

The answers are more compact in the limit  $|k|\gg 1$, corresponding to the
``deep water'', when $\hat R\to \hat H$, $\hat T\to -\hat H$, with $\hat H$ 
being the Hilbert operator: $\hat H=i\,\mbox{sign\,}\hat k$. 
In this case  ${\cal K}^{(1)}_{[b]}\to 0$, and therefore
\begin{equation}\label{Hamiltonian_deep}
{\cal K}_{deep}\approx-\frac{1}{2}\int\psi\,\hat H\psi_u\, du\,dq 
+{\cal F}\{\Psi,\overline{\Psi},Z,\overline{Z}\}.
\end{equation}
After appropriate rescaling of the variable $u$, we can write
$$
Z=u+(i-\hat H)Y(u,q,t),\qquad Z_u=1+(i-\hat H)Y_u.
$$
The equations of motion for quasi-plane waves on the deep water look as follows:
\begin{equation}\label{kinematic_deep}
Z_t=Z_u(\hat H-i)\left[{[\hat H\psi_u-(\delta{\cal F}/\delta\psi)]}/
{|Z_u|^2}\right],
\end{equation}
\begin{eqnarray}
\psi_t+g\,Y&=&
\,\psi_u\hat H\left[{[\hat H\psi_u-(\delta{\cal F}/
\delta\psi)]}/{|Z_u|^2}\right]\nonumber\\
&+&{\hat H\left[\psi_u [\hat H\psi_u-(\delta{\cal F}/
{\delta\psi})]\right]}/{|Z_u|^2}\nonumber\\
\label{Bernoulli_deep}
&-&{2\,\mbox{Re}\left((\hat H+i)
[Z_u({\delta{\cal F}}/{\delta Z})]\right)}/{|Z_u|^2},
\end{eqnarray}
where $(\delta{\cal F}/\delta\psi)=2\,\mbox{Re\,}\left[
(1-i\hat H)(\delta{\cal F}/{\delta\Psi})\right]$.

\section{Summary}

Thus, now we have nonlinear evolution equations for weakly 3D steep 
water waves, as for deep water case, as for arbitrary quasi-1D bottom 
profile. The obtained equations are intended to describe, for example, 
the sudden formation of giant waves in open sea, 
as well as overturning waves on a beach.
The following step should be development of an efficient numerical method 
for simulation these equations.

%%%%%%%%%%%%%%%%%%%%%%%%%%%%%%%%%%%%%%%%%%%%%%%%%%%%%%%%%%%%%%%%%%%%%%

\end{document}